\documentclass[aps,prl,showpacs,amsmath,amssymb,graphicz,epsf,twocolumn]{revtex4}
\usepackage{graphicx}
\usepackage{epsf}
\usepackage{graphicx}
\begin{document}
\title{Anisotropic Transport Properties of Ferromagnetic-Superconducting 
Bilayers.}
\author{M. Amin Kayali$^{1,*}$, Valery L. Pokrovsky$^{1,2}$}
\affiliation{$^1$Department of Physics, Texas A \& M
University,College Station,
Texas 77843-4242\\
$^2$ Landau Institute for Theoretical Physics, Moscow, Russia}

\begin{abstract}
We study the transport properties of vortex matter in a 
superconducting thin film separated by a thin insulator layer from 
a ferromagnetic layer. We assume an alternating stripe structure 
for both FM and SC layers as found in \cite{ELPV}. We calculate 
the periodic pinning force in the stripe structure resulting 
from a highly inhomogeneous distribution of the vortices and 
antivortices. We show that the transport in SC-FM bilayer is 
highly anisotropic. In the absence of random pinning it displays 
a finite resistance for the current perpendicular to stripe and is 
superconducting for the current parallel to stripes. The average 
vortex velocity, electric field due to the vortex motion, 
Josephson frequency and higher harmonics of the vortex oscillatory 
motion are calculated.
\\
\end{abstract}
\pacs{74.60.Ge, 74.76.-w, 74.25.Ha, 74.25.Dw}
\maketitle 
The interest in Heterogeneous ferromagnetic-superconducting systems 
has grown rapidly in recent years. This interest stems not only from their 
possible technological applications but also from new physical phenomena 
arising from the interaction between two order parameters. 
Typically such a system consists of a superconductor (SC) placed 
in close proximity with a periodic ferromagnetic structure (FS) 
such as an array of ferromagnetic dots or holes. 
The two systems are separated by an infinitely
thin layer of insulator oxide that guarantees the suppression of
proximity effects. It was demonestrated experimentally \cite{morgan}-\cite{mosch} 
that the interaction between the superconductor and the ferromagnet may lead to  
formation of superconducting vortices interacting with the FM film. Theoretical studies
of such systems have been done in \cite{POLY1}-\cite{ELPV} 

Recently, Erdin {\it{et.al.}}\cite{ELPV} studied the equilibrium 
structure of a FM-SC bilayer (FSB). They have proved that it represents a
two-dimensional periodic stripe domain structure consisting of two
equivalent sub-lattices, in which both the  magnetization $m_z(\bf{r})$
and the vortex density $n_v (\bf{r})$ alternate. Thus, they predicted 
spontaneous violation of the translational and rotational symmetry
in the bilayer. In this article we study the transport
properties of the FSB. They are associated with the driving force acting
on the vortex lattice from an external electric current. 
We show that the FSB exhibits strong anisotropy of the transport 
properties: the bilayer may be superconducting for the current parallel to 
the domain walls and resistive when the current is perpendicular to them.
 
Periodic pinning forces in the direction parallel to the stripes do not appear 
in continuously distributed vortices, their reappearance is associated with 
the discreteness of the vortex lattice. Therefore, we need to modify the theory 
\cite{ELPV} to incorporate the discreteness effects.  Let us 
assume that the saturation magnetization per unit area of the FM film is 
$m$ and its width is $L$. The energy necessary to create a 
single Pearl vortex \cite{pearl} in the superconductor is 
$\epsilon_{v0}=\epsilon_0 \ln(\frac{\lambda}{\xi})$, 
with $\epsilon_0=\frac{\phi_0^2}{16\pi^2\lambda}$ 
where $\phi_0$ is the flux quantum, $\lambda=\frac{\lambda_L^2}{d_s}$ 
is the effective penetration depth \cite{abrik}, $\lambda_L=
\sqrt{\frac{m _ec^2}{4\pi n_s e^2}}$ is the London 
penetration depth, $d_s$ is the thickness of the superconducting 
layer and $\xi$ is its coherence length. It was shown in 
\cite{ELPV} that the interaction between 
the superconducting vortices and the magnetization in the stripe 
structure renormalizes the single-vortex energy to the value 
$\tilde {\epsilon_v}=\epsilon_{v0}- m \phi_0$ which must be negative 
to allow development of the stripes. The density of the superconducting 
vortices increases when approaching the domain walls and can be 
expressed as $n_v(x)=\frac{\pi \tilde{m}}
{L\phi_0}\frac{1}{\sin(\frac{\pi x}{L})}$ where $\tilde{m}=m 
-\frac{\epsilon_{v0}}{\phi_0}$ is the renormalized magnetization of 
the FM stripe. The vortices spontaneously appear in the superconductor. 
We assume that the vortices inside one stripe are arranged in parallel 
chains. Each chain is periodic with the same lattice constant 
$b$ along the chain, whereas the distance between $k$-th and $(k+1)$-th 
chain $a_k$ depends on $k$. The correspondence between this discrete 
arrangement and continuous approximation \cite{ELPV} is established by 
the requirement that the local vortex density $n_v (x_k)$ 
calculated in \cite{ELPV} must be equal to $(ba_k )^{-1}$. 
The coordinate $x_k$ is determined in terms of $a_k$ as a sum: 
$x_k =\sum_{k^\prime =0}^{k-1} a_{k^\prime}$. For definiteness we choose the 
origin in the center of the stripe. We assume that the total 
number of the vortex chains $2N$ in a stripe is large. Then some of 
them are located very close to the domain walls. Let us remind that, 
in the continuous approximation \cite{ELPV} $n_v= \frac{\pi \tilde{m}}
{L\phi_0}\frac{1}{\sin(\frac{\pi x}{L})}$, where $L$ is the domain width. 
Considering the nearest to the domain wall vortex chain 
(with the number, N), we put $n_v(x_N)=\frac{1}{b a_N}$. 
On the other hand $x_N =L-a_N$. 
Since $\frac{a_N}{L} \ll 1$, we find: $b=\frac{\phi_0}{\tilde{m}}$. 
The total number of chains in a stripe is $2N$, where
$N=b\int_0^{L-\lambda} n_v (x)dx =\frac{1}{2}\ln(\frac{L}{\lambda})$. 
We cut the integration (and summation) 
at a distance $\sim \lambda$ from the domain wall where the 
continuous approximation breaks. Thus the minimum value 
of $a$ is $\lambda$.
When transport current passes through the superconducting film, the vortices 
start to move. To simplify the problem we assume that all vortices in each 
stripe move together as well as all antivortices in the 
neighboring stripe do.
\begin{figure}[t]
\centering
 \includegraphics[angle=0,width=3.0in,totalheight=2.0in]{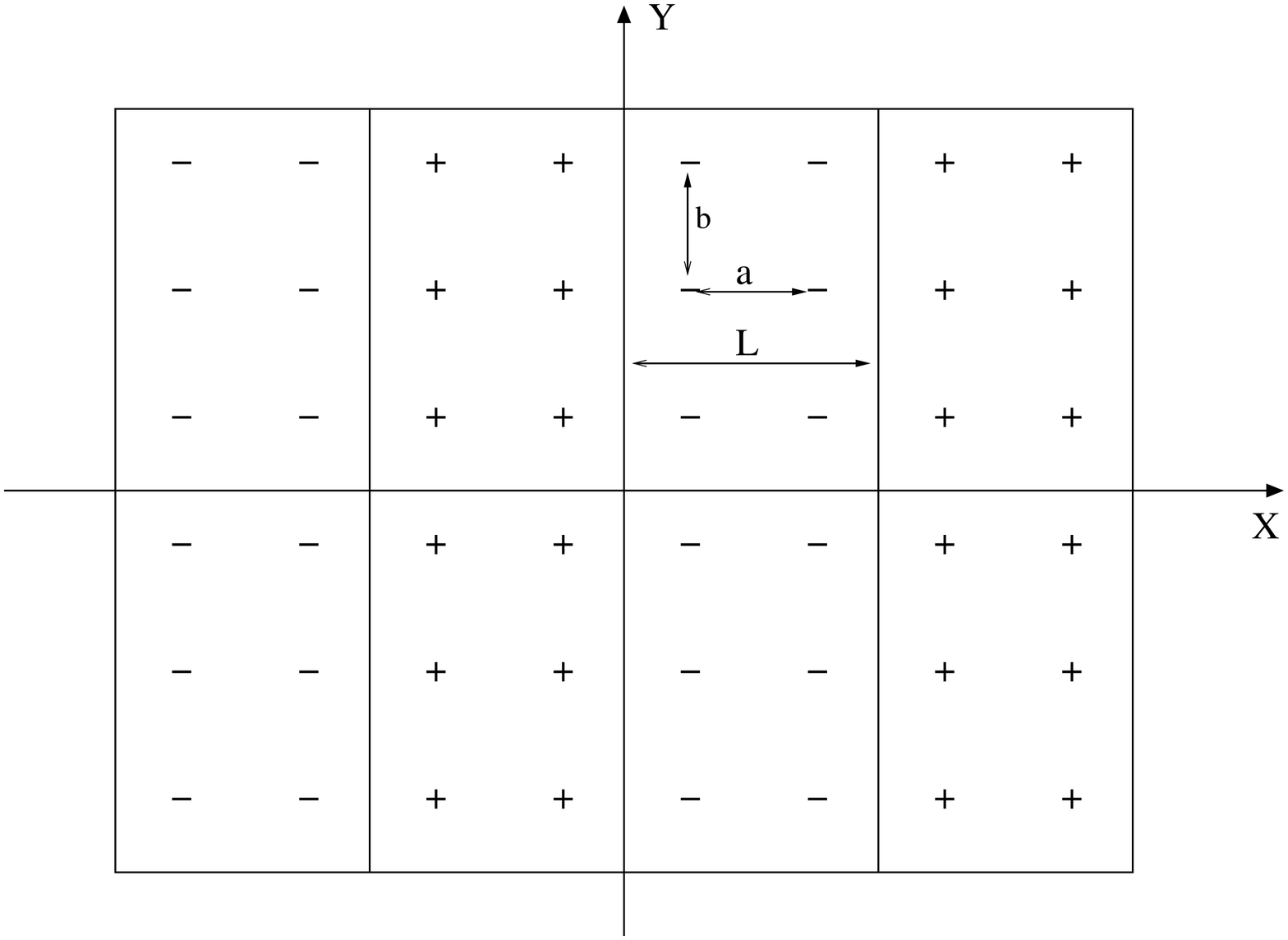}
 \caption{Schematical vortex distribution in the FM-SC bilayer. 
   The sign $\pm$ refers to the vorticity of the trapped flux.}
\label{fig1}
\end{figure}
We denote their positions ${\bf{r}}_{+}=(x_{+},y_{+})$ and 
${\bf{r}}_{-}=(x_{-},y_{-})$, respectively. Forces acting on a moving vortex 
are the Magnus force, the viscous force and the periodic 
pinning force.The Magnus force is ${\bf{f}}_m=\pi n_s \hbar d_s (\bf{v}_s 
-\dot{\bf{r}})\times \hat{z}$, where $n_s$ is the superconducting electron 
density, $v_s$ is the velocity of the superconducting electron and 
$\dot{\bf{r}}$ is the vortex velocity. The viscous (friction) force 
is ${\bf{f}}_f=-\eta \dot{\bf{r}}$ where $\eta= \frac
{\phi_0 H_{c2}d_s}{\rho_n c^2}$ is the Bardeen-Stephen drag coefficient 
\cite{bardeen}, $H_{c2}$ is the upper critical magnetic field, 
$\rho_n$ is the resistivity of the superconducting sample in the normal 
state, and $c$ is the speed of light. The periodic pinning forces are 
due to the interaction of the vortex with the pinning 
centers and the domain walls. In the FM-SC bilayer the pinning force is 
due to the interaction of the domain walls with the vortices and antivortices 
and the vortex-vortex interaction $U_{vv}$ given by:
\begin{eqnarray}
U_{vv}=\frac{1}{2}\int \int n_v({\bf{r}}) V({\bf{r}}-{\bf{r^{\prime}}}) 
n_v ({\bf{r^{\prime}}}) d^2{\bf{r}} d^2{\bf{r^{\prime}}}
\end{eqnarray}
wher $V({\bf{r}}-{\bf{r^{\prime}}})$ is the pair interaction between 
vortex located at ${\bf{r}}$ and another at ${\bf{r^{\prime}}}$. When 
$|{\bf{r}}- {\bf{r^{\prime}}}|\gg \lambda$ the pair interaction can be 
written as $ V({\bf{r}}-{\bf{r^{\prime}}})=\frac{\phi_0^2}{4\pi^2}
\frac{1}{|{\bf{r}}- {\bf{r^{\prime}}}|}$ \cite{degen}.
The interaction energy between two parallel chains located at $x_l$ 
and $x_{l^\prime}$ and vertically shifted with respect to each other 
by an interval $b s$ ($s\ll 0$) reads: 
\begin{eqnarray}
U(x_l ,x_{l^{\prime}},s) =\sum_{n,m=1}^{N_0}
\frac{\frac{\phi_0^2}{8\pi^2} }{\sqrt{(x_l -x_{l^\prime})^2 
+(n-m-s)^2 b^2}}
\label{poten}
\end{eqnarray}
where $N_0$ is the number of vortices or antivortices in a single chain. 
For infinite chains ($N_0 \rightarrow \infty$) Eq.(\ref{poten}) 
can be rewritten as
\begin{eqnarray}
U(x_l ,x_{l^{\prime}},s) =\sum_{k=-\infty}^{\infty}
\frac{\frac{N_0 \phi_0^2}{8\pi^2} }{\sqrt{(x_l -x_{l^\prime})^2 
+(k-s)^2 b^2}}
\label{poten1}
\end{eqnarray}
The sum in (\ref{poten1}) can be calculated using Poisson summation 
formula \cite{Grad}. Since the force is zero in the continuous 
approximation, it is possible to retain the lowest non-zero 
harmonic in the Poisson summation. Thus, we arrive at the following 
interaction energy of two chains: 
\begin{eqnarray}
U(x_l ,x_{l^{\prime}},s)= \frac{N_0 \phi_0^2}{4\pi^2 b} \cos(2\pi s) 
\chi_{l l^\prime} 
\label{poten2}
\end{eqnarray}
where $\chi_{l l^\prime}=e^{-2\pi \frac{|x_l -x_{l^\prime}|}{b}}$.
The distance between two chains $|x_l -x_{l^\prime}|$ is larger or equals 
$\lambda$ hence $\chi_{l l^\prime} \sim \chi=e^{-\frac{2\pi\lambda}{b}}$. 
Typical value of $\chi_{l l^\prime}$ is $e^{-\frac{\delta_m}{4\pi}}$  where 
$\delta_m=\frac{m \phi_0}{\epsilon_0} =g S\frac{n_m d_m}{n_s d_s}$, 
with $g$ is Lande factor, $S$ is the ferromagnet elementary spin, $n_m$ 
and $d_m$ are the electrons density and thickness of the magnetic 
film respectively. 

We conclude that the amplitude of the periodic potential for displacements
parallel to the domains is exponentially small in units of $\epsilon_0$, 
near the transition temperature. Relative displacements in perpendicular
direction have energy barrier $\sim \epsilon_0$ even in continuous 
approximation. We model the restoring forces by simple sines dependencies
$f_x =f_\bot \sin(\frac{2\pi}{a}(x_{+} -x_{-}))$,  
$f_y= f_{||} \sin(\frac{2\pi}{b} (y_{+}-y_{-}))$, where
$f_\bot \sim \frac{\epsilon_0}{a}$ and 
$f_{||}\sim \frac{\epsilon_0}{b}e^{-\frac{\delta_m}{4\pi}} \ll f_\bot$.
 
When the supercurrent is perpendicular to domains, the equations of motion 
for a vortex and antivortex 
\begin{eqnarray}
\eta \dot{y}_{+} &=& F -\frac{F}{v_s}\dot{x}_{+} -f_{||}
\sin(\frac{2\pi}{b}(y_{+} -y_{-})) \label{eq1}\\
\eta \dot{x}_{+} &=& \frac{F}{v_s}\dot{y}_{+} +f_\bot
\sin(\frac{2\pi}{a}(x_{+}-x_{-}))\label{eq2}\\
\eta \dot{y}_{-} &=& -F +\frac{F}{v_s}\dot{x}_{+} +f_{||}
\sin(\frac{2\pi}{b}(y_{+} -y_{-})) \label{eq3}\\
\eta \dot{x}_{-} &=& -\frac{F}{v_s}\dot{y}_{-}
-f_\bot \sin(\frac{2\pi}{a}(x_{+}-x_{-})) 
\label{eq4}
\end{eqnarray}
where $F=\pi n_s\hbar d_s v_s$, $f_\bot=\frac{\epsilon_0}{a}$ and 
$f_{||}=\frac{\chi \epsilon_0}{b}$. 
If the current is smaller than a critical value, Eq.(\ref{eq1}-\ref{eq4}) 
accept a static solution
\begin{eqnarray}
x_{+}&=&x_{-}= \frac{Fb}{4\pi \eta v_s}\arcsin(\frac{F}{f_{||}}) 
\label{xstatsol}\\
y_{+}&=&-y_{-}= \frac{b}{4\pi} \arcsin(\frac{F}{f_{||}}) 
\label{ystatsol}
\end{eqnarray}
It is valid at $F\leq f_{||} =\frac{\chi \epsilon_0}{b}$. For $F>f_{||}$ 
or equivalently, if the current is larger than its critical value, 
vortices and antivortices start to move. The solution of 
Eq.(\ref{eq1}-\ref{eq4}) for $F>F_{c}$ reads:
\noindent
\begin{eqnarray}
x_{+}-x_{-}&=&0\label{eq7}\\
x_{+}+x_{-}&=&\frac{F}{\eta v_s} (y_{+}-y_{-})\label{eq8}\\
y_{+}-y_{-}&=&\frac{b}{\pi}
\arctan(\frac{f_{||}}{F} +\sqrt{1 -\frac{f_{||}^2}
{F^2}}\tan(\omega_0^{\bot} t))
\label{eq9}\\
y_{+}+y_{-}&=&0\label{eq10}
\end{eqnarray}
\begin{figure}[t]
\centering
 \includegraphics[angle=0,width=3.5in,totalheight=2.0in]{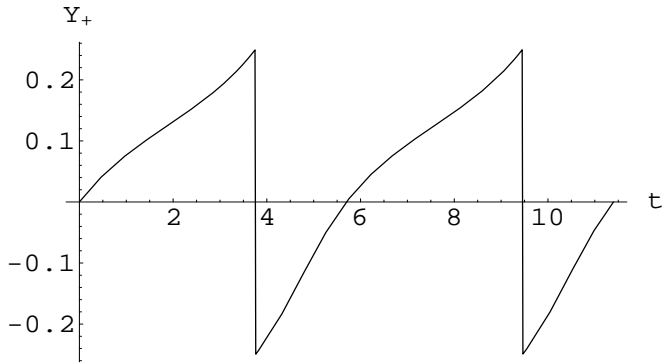}
  \caption{The vortex displacement as a function of time in the overcritical regime. 
Time is measured in units of $\frac{1}{\omega_0^{\bot}}$ and $y_{+}$ in 
units of $b$ and $\chi=10^{-4}$.} 
\label{figure1}
\end{figure}

\noindent
where $\omega_0^{\bot}=\frac{2\pi\eta v_s^2 \sqrt{b^2 F^2 
-\chi^2 \epsilon_0^2}} {b^2(F^2 +\eta^2 v_s^2)}$ is the Josephson frequency. 
Thus the vortices and antivortices acquire the same velocity 
components $v_{+x}=v_{-x}$ in the direction of the current and 
opposite velocity components $v_{+y}=-v_{-y}$ in the direction 
perpendicular to the current. The domain walls do not interfere 
such a motion if they move in the direction of the current with 
the same velocity $v_{dw}=v_{+x}=v_{-x}$ as vortices and antivortices. 
Such a motion is a Goldstone mode.
The solution (\ref{eq7}-\ref{eq9}) displays an oscillatory motion 
of the vortices and antivortices in the direction parallel to the 
domain walls, in addition to their motion together with the domain walls 
along the direction of the current. Higher harmonics of the vortex 
motion can be calculated analytically. 
The distribution of vortices (antivortices) is inhomogeneous
in the direction perpendicular to the domains. The local electric field 
$\bf{E}$ due to the vortex motion is related to
its time-average velocity $<\bf{v}_+>$ as
${\bf{E}}= -\frac{q_v \phi_0}{c}n_v ({\bf{r}}) (<{\bf{v}}_{+}> 
\times \hat{{\bf{z}}})$ \cite{larkin}. Therefore, the local field 
produced by vortices in the direction parallel to the domains is 
equal but opposite in sign to the one produced by antivortices, while 
the local field produced by vortices and antivortices in the direction 
perpendicular to the domains has both equal magnitude and sign.
The time average of the vortex (antivortex) velocity over a period 
$T=\frac{2 \pi}{\omega_0^{\bot}}$ is 
\begin{eqnarray}
<{\bf{v}}_{+y}> &=&\pm\frac{\eta \sqrt{F^2 -f_{||}^2}}
{(\eta^2 +\frac{F^2}{v_s^2})}\label{perpvel}\\
<{\bf{v}}_{+x}> &=&\frac{F<v_{+y}>}{\eta v_s} \label{paralvel}
\end{eqnarray}
The time-averaged local field components are
\begin{eqnarray}
E_x=-\frac{\eta \tilde{m}}{a c} \frac{
\sqrt{F^2 -f_{||}^2}}{(\frac{F^2}{v_s^2}+\eta^2)}\label{locex}\\
E_y=\mp \frac{\tilde{mF}}{a v_sc} \frac{
\sqrt{F^2 -f_{||}^2}}{(\frac{F^2}{v_s^2}+\eta^2)}\label{locey}
\end{eqnarray} 
The upper sign in Eq.(\ref{perpvel}) and Eq.(\ref{locey}) refer to 
the vortices velocity and produced field along the domain while the 
lower sign refer for those due to antivortices. Non-zero average electric field due 
to all vortices and antivortices in the FSB appears only in the direction perpendicular 
to the domains.   
The critical current $J_c$ is related to $F_c$ as $J_c=\frac{c}{\phi_0 d_s} F_c$.
Plugging $F_c=\frac{\epsilon_0 m \chi}{\phi_0}$ into the expression for $J_c$ and 
accepting $\chi=10^{-4} - 10^{-2}$, $b=10^{-4}-10^{-5} cm$, and 
$n_s=10^{22} cm^{-3}$ we find: $J_c\sim 10^3 -10^5 \frac{A}{cm^2}$  

When the current flows parallel to the stripes, the FM domain walls 
stay at rest while vortices and antivortices move both parallel and 
perpendicular to the domains. The solution of equations of 
motion for vortices and antivortices shows that they move opposite to one another 
both in $x$ and $y$ directions. Their motion along $x$ is oscillatory with 
fundamental frequency $\omega_0^{||}=\frac{2\pi\eta v_s^2 
\sqrt{a^2 F^2 -\epsilon_0^2}} {a^2(F^2 +\eta^2 v_s^2)} $. The 
motion of vortices and antivortices in the parallel direction proceeds 
until the distance between them becomes half lattice spacing $\frac{b}{2}$. 
Once the vertical shift between the vortices and antivortices reaches 
$\frac{b}{2}$, their motion freezes. The critical current in this case is 
$J_c=\frac{n_s \mu_B}{2 a}$, 
the lattice spacing $a$ is of the order of $\lambda \sim 10^{-5}-10^{-4} cm$, 
hence the critical current $J_c$ is of the order $10^{7}-10^{8} \frac{A}{cm^2}$, 
which is at least  $10^2$ times larger than the critical current for parallel 
current. Therefore, the system may be superconducting for the current parallel 
to the stripes and exhibit finite resistance for perpendicular current. 
The difference in the critical currents for parallel and perpendicular 
directions is due to the exponential factor $\chi$ which is small if $b \ll \lambda$. 
The anisotropy is pronounced when $\delta_m$ is large which can be achieved by 
using thicker FM layers and decreasing the density of the 
superconducting electrons. $\delta_m$ is temperature dependent and eventually 
decreases when temperature decreases starting from $T_s$. However, at the 
temperature of vortex disappearance $T_v < T_s$ the value $\tilde{m}$ turns into zero
and $\chi$ again becomes exponentially small. Thus anisotropy has a minimum between 
$T_v$ and $T_s$.

Kopnin and Vinokur \cite{KopVin} considered a collection of superconducting grains with 
the washboard pinning potentialas a model of random pinning. They obtained a similar result for
vortex sliding in external magnetic field with a supercurrent applied. In contrast to their 
work (they considered vortices only), we consider vortices and antivortices in the field 
of periodic pinning and completely neglect the random pinning.    

Let us discuss briefly how the magnetic field generated by supercurrent 
changes our result. In \cite{AKPS} it was shown that at sufficiently small critical 
magnetic field the domains vanish. Therefore, in general, 
magnetic field suppresses both the anisotropy and periodic pinning 
at a critical field for which domains disappear. At such 
critical field only random pinning prevails. However, the total per unit length current is 
proportional to the thickness of the SC film and can be kept small.

In conclusion, we studied the transport properties of the 
FM-SC bilayer in a state with stripe domains of
alternating magnetization and vorticity. We showed that, in the absence of a 
driving force, the vortices and antivortices lines themselves 
up in straight chains configuration. We showed that 
the force between two chains of vortices falls off exponentially 
as a function of the distance separating the chains. We argued that 
the exponential decay of the  pinning force in the direction 
parallel to the domains drops faster in the vicinity of the 
superconducting transition temperature $T_s$ and vortex disappearence temperature $T_v$. 
We solved the equations of motion for vortices and antivortices in two cases for the driving 
current direction, parallel or perpendicular to the domains. 
We calculated the critical current for both cases and found that 
its value for the parallel current is much higher than for 
perpendicular one. Our most important result is a strong anisotropy of the 
critical current. We expect the ratio of the parallel to perpendicular critical
current is in the range $10^2 \div 10^4$ close to the superconducting 
transition temperature $T_s$ and to the vortex disappearance temperature $T_v$.
The anisotropy decreases rapidly when the temperature goes from the ends of this
interval reaching its minimum somewhere inside it.
This anisotropic transport behavior could serve as a diagnostic tool 
to discover spontaneous topological structures in magnetic-superconducting systems.

This work was supported by the NSF grants DMR 0103455 and DMR 0072115, 
DOE grant DE-FG03-96ER45598, and Telecommunication and Information Task 
Force at Texas A \& M University.

\end{document}